\documentclass{JINST}

\usepackage{textcomp}

\title{Design and operation of a cryogenic charge-integrating preamplifier 
for the MuSun experiment}

\author{
R.A.~Ryan$^a$,
F.~Wauters$^a$,
F.E.~Gray$^{b,a}$\thanks{Corresponding author.} ,
P.~Kammel$^a$,
A.~Nadtochy$^c$,
D.~Peterson$^a$,
T.~van~Wechel$^a$,
E.~Gross$^b$,
M.~Gubanich$^b$,
L.~Kochenda$^c$,
P.~Kravtsov$^c$,
D.~Orozco$^b$,
R.~Osofsky$^a$,
M.H.~Murray$^a$,
G.E.~Petrov$^c$,
J.D.~Phillips$^d$,
J.~Stroud$^b$,
V.~Trofimov$^c$,
A.~Vasilyev$^c$
and M.~Vznuzdaev$^c$ \\
\llap{$^a$} Department of Physics, University of Washington \\ Seattle, WA 98195, USA \\
\llap{$^b$} Department of Physics and Computational Science, Regis University \\ Denver, CO 80221, USA \\
\llap{$^c$} High Energy Physics Division, Petersburg Nuclear Physics Institute \\ Gatchina 188350, Russia \\
\llap{$^d$} Department of Physics, Boston University \\ Boston, MA 02215, USA \\
  E-mail: \email{fgray@regis.edu}}

\abstract {
The central detector in the MuSun experiment is a pad-plane 
time projection ionization chamber that operates without gas amplification 
in deuterium at 31~K; it is used to measure the rate of the muon capture 
process $\mu^- + d \rightarrow n + n + \nu_\mu$.  A new charge-sensitive 
preamplifier, operated at 140~K, has been developed for this detector.  
It achieved a resolution of 4.5~keV(D$_2$) or 120~$e^-$ RMS with zero detector 
capacitance at 1.1~{\textmu}s integration time in laboratory tests.  In the 
experimental environment, the electronic resolution 
is 10~keV(D$_2$) or 250~$e^-$ RMS at a 0.5~{\textmu}s integration time.
The excellent energy resolution of this amplifier has enabled discrimination 
between signals from muon-catalyzed fusion and muon capture on chemical 
impurities, which will precisely determine systematic 
corrections due to these processes.  It is also expected to improve the 
muon tracking and determination of the stopping location.
}

\keywords{Front-end~electronics~for~detector~readout;
Analogue~electronic~circuits;
Time~projection~chambers}

\begin{document}

\section{Introduction}

The MuSun experiment~\cite{Andreev:2010wd} at the Paul Scherrer Institute (PSI)
will provide a high-precision measurement of the nuclear capture rate of 
negative muons by deuterons ($\mu^- + d \rightarrow n + n + \nu_\mu$), 
a fundamental reaction that calibrates effective field theories of two-nucleon 
weak interactions~\cite{doi:10.1146/annurev-nucl-100809-131946}.
Muons are stopped within a pad-plane time projection chamber (TPC)
where deuterium gas is both the target material and the active
medium of the detector.  Each muon is tracked by the TPC to its stopping point,
and only those muons that stop in the pure deuterium, away from the chamber 
walls, are accepted for later analysis.  This technique is based in part on 
the experience derived from the TPC in the MuCap experiment~\cite{MuCapTPC},
which was the predecessor of MuSun.

\begin{figure}
\begin{center}
(a)
\includegraphics[width=0.9\textwidth]{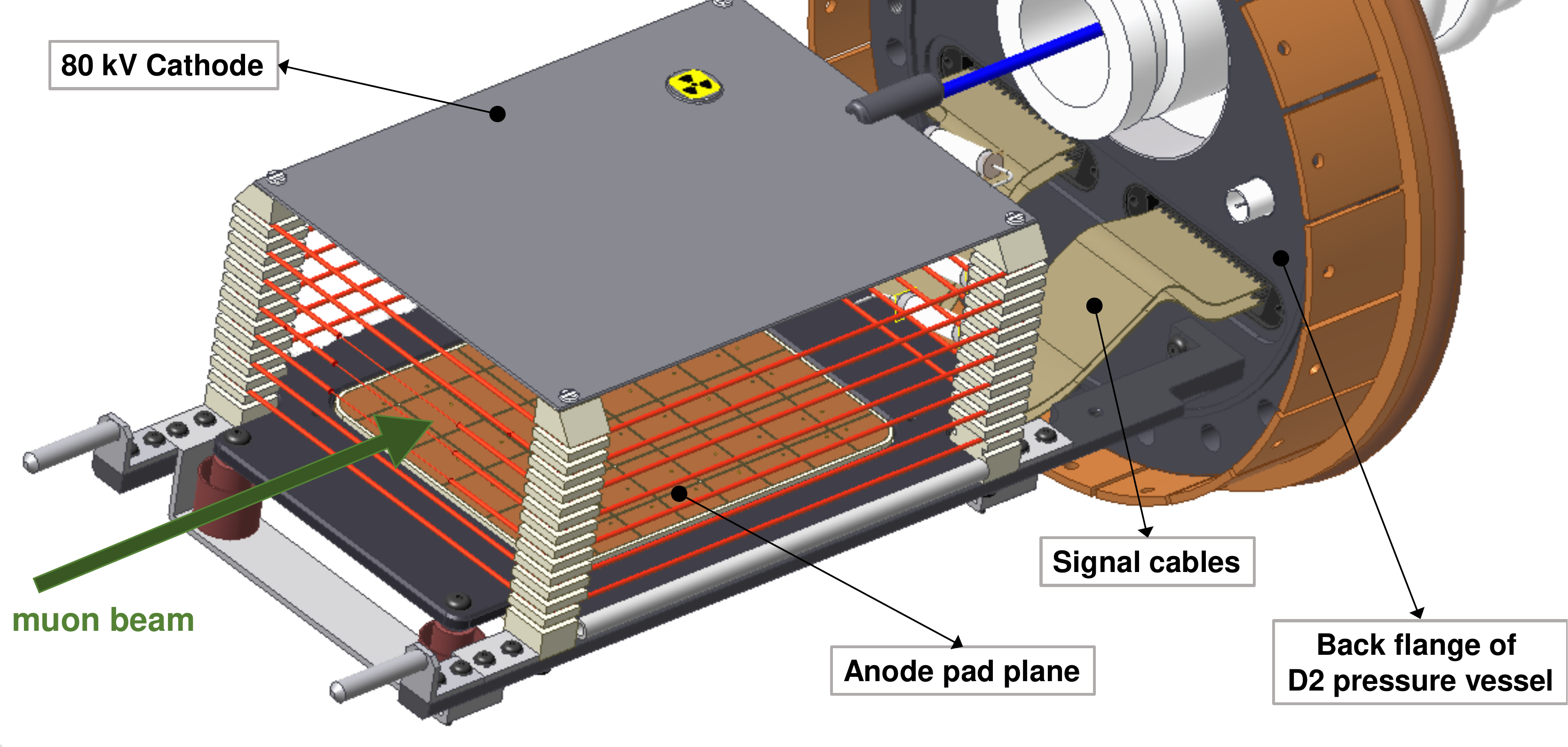}

\vspace{1.5cm}

(b) 
\includegraphics[width=0.9\textwidth]{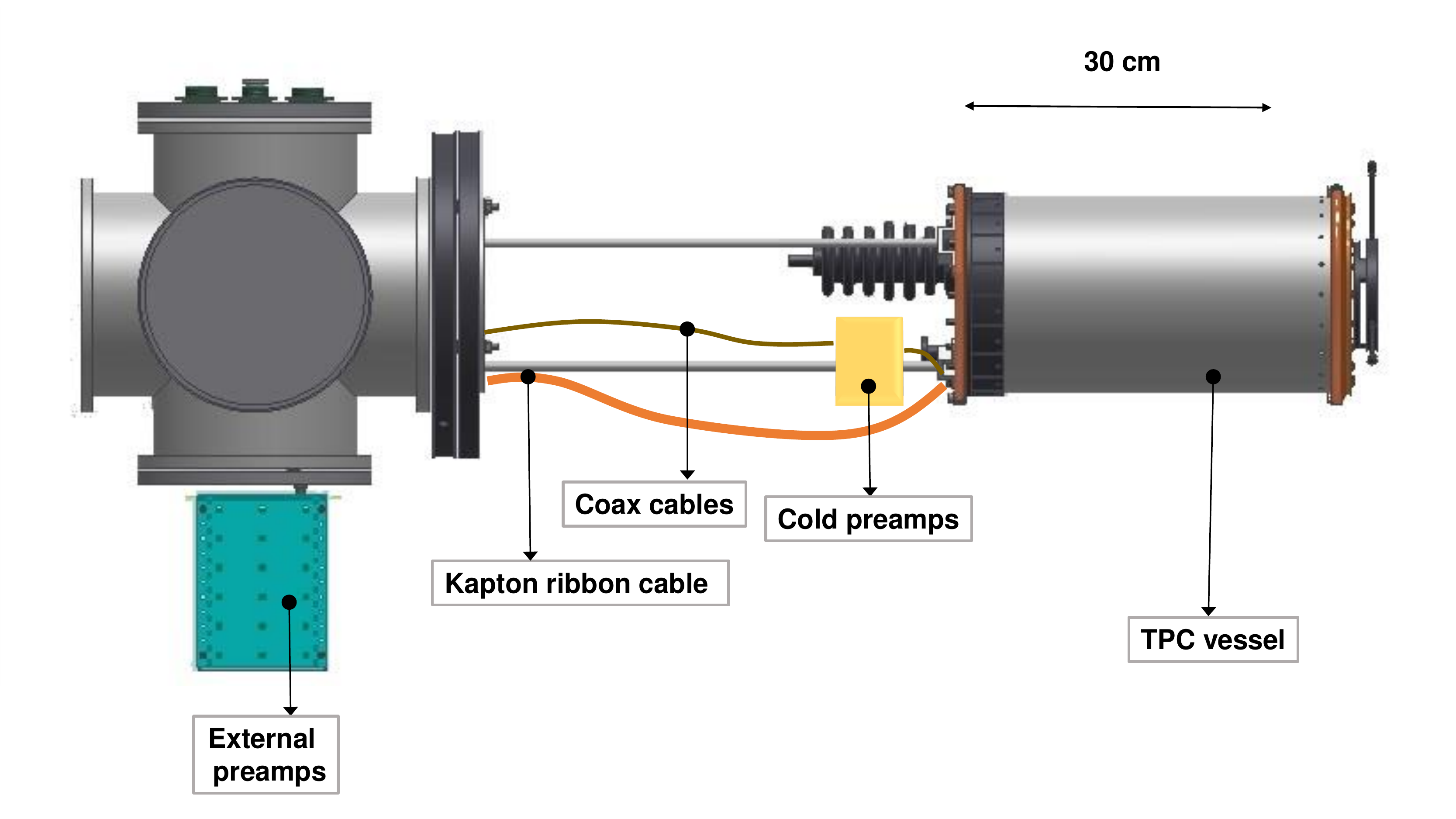}

\end{center}
\caption{(a) View inside TPC to illustrate pad plane structure. 
(b) Simplified overview of the MuSun TPC apparatus, 
showing the locations of the old external room temperature preamplifiers and 
the new cold preamplifiers.}
\label{fig:overall}
\end{figure}

The TPC's 48 pads, placed at a pitch of 16~mm~$\times$18~mm, are arranged 
in a 6$\times$8 array, as shown in Figure~\ref{fig:overall}.
The TPC detects not only muon stops but also 
the products of muon-catalyzed fusion, as well as the recoil nuclei
following muon capture on chemical impurities in the gas.   
Table~\ref{tab:processes}
gives the scale of characteristic energy deposits from each of these processes.\footnote{Fusion occurs from $d d \mu$ molecular states and leads to monoenergetic
recoils.  Muon capture on impurities produces continuous recoil energy 
distributions which must be determined experimentally.  The estimates for the
characteristic energy scale of these processes are based on the simplest 
case that an excited daughter nucleus recoils against the neutrino.}
We are motivated to improve the energy resolution of the chamber to its 
physical limit in order to be able to separate capture on nitrogen and oxygen 
impurities from fusion to the $^3$He~+~$n$ final state.  In this way, the TPC 
itself is used to monitor the purity of the deuterium. 
The improved resolution also provides a more precise determination of
the muon stopping point, which is determined by comparing the energy 
deposited in the last and next-to-last pads in the track.
It may even provide a method to identify the muon stopping
point based on a definition that does not rely on the final pad at all, 
instead using an extrapolation of the much smaller energy deposits
in upstream pads to project the location of the Bragg peak.
This method would reduce the sensitivity of the measurement to the 
muon-catalyzed fusion pulses, which appear near the muon stopping point.
\begin{table}
\begin{center}
\begin{tabular}{lrr}
\hline
Process & Energy scale (MeV) & Visible energy scale (MeV) \\
\hline
$\mu$ stop & 0.2 to 0.8 per pad & 0.2 to 0.8 per pad \\
$d d \mu \rightarrow \mu^-$ + ${}^3$He + $n$ & 0.82 & 0.35 \\
$d d \mu \rightarrow \mu {}^3$He + $n$ & 0.80  &  0.53 \\
$d d \mu \rightarrow \mu^-$ + $t$ + $p$ & 1.01 ($t$) + 3.02 ($p$) & 0.8 ($t$) + 3.0 ($p$)\\
$\mu {}^{14}$N $\rightarrow$ ${}^{14}$C\** + $\nu_\mu$ & $\sim$0.30 & $\sim$0.14 \\
$\mu {}^{16}$O $\rightarrow$ ${}^{16}$N\** + $\nu_\mu$ & $\sim$0.26 & $\sim$0.14 \\
\hline
\end{tabular}
\end{center}
\caption{Physical processes that deposit significant energy in the TPC.
The visible energy is reduced from the physical energy deposition primarily
because of recombination of electron/ion pairs, which increases with the 
density of space charge.  The visible energy scale therefore depends strongly 
on the gas density; it also depends on the transparency of the Frisch grid.}
\label{tab:processes}
\end{table}

In order to minimize the number of these muon-catalyzed fusion pulses,
the deuterium is cooled to 31~K.   This temperature is just above 
the liquefaction point for its pressure of 5.1~bar, giving a density of
6.4\% of liquid hydrogen.  The purity of the deuterium must be maintained 
at the level of 1 part per billion (ppb), so the 
materials that contact the gas were selected for minimal outgassing, 
and the gas is continuously circulated through a cryogenic absorber~\cite{Ganzha:2007uk}.
Because the capture rate scales approximately as $Z^4$~\cite{Measday:2001yr},
contamination with nitrogen or oxygen increases the measured capture rate 
by about 2~s$^{-1}$ per ppb. Consequently, even a very small impurity 
concentration will require a significant correction, in comparison to 
the experiment's target precision of $\pm$6~s$^{-1}$. 

The drift field in the TPC is generated by a cathode plane, 72.4~mm above 
the pads, that is held at -80~kV; this gives a maximum drift time of 13.5~{\textmu}s
from the top of the chamber.  A Frisch grid~\cite{Bunemann}, at -3.5~kV, 
is 1.5~mm above the pads, which are at ground potential.  There is no gas 
amplification in this arrangement; the Frisch grid improves the time resolution 
and eliminates the dependence of the pulse amplitude on the position
of the ionization.   

\begin{figure}[b]
\includegraphics[width=\textwidth]{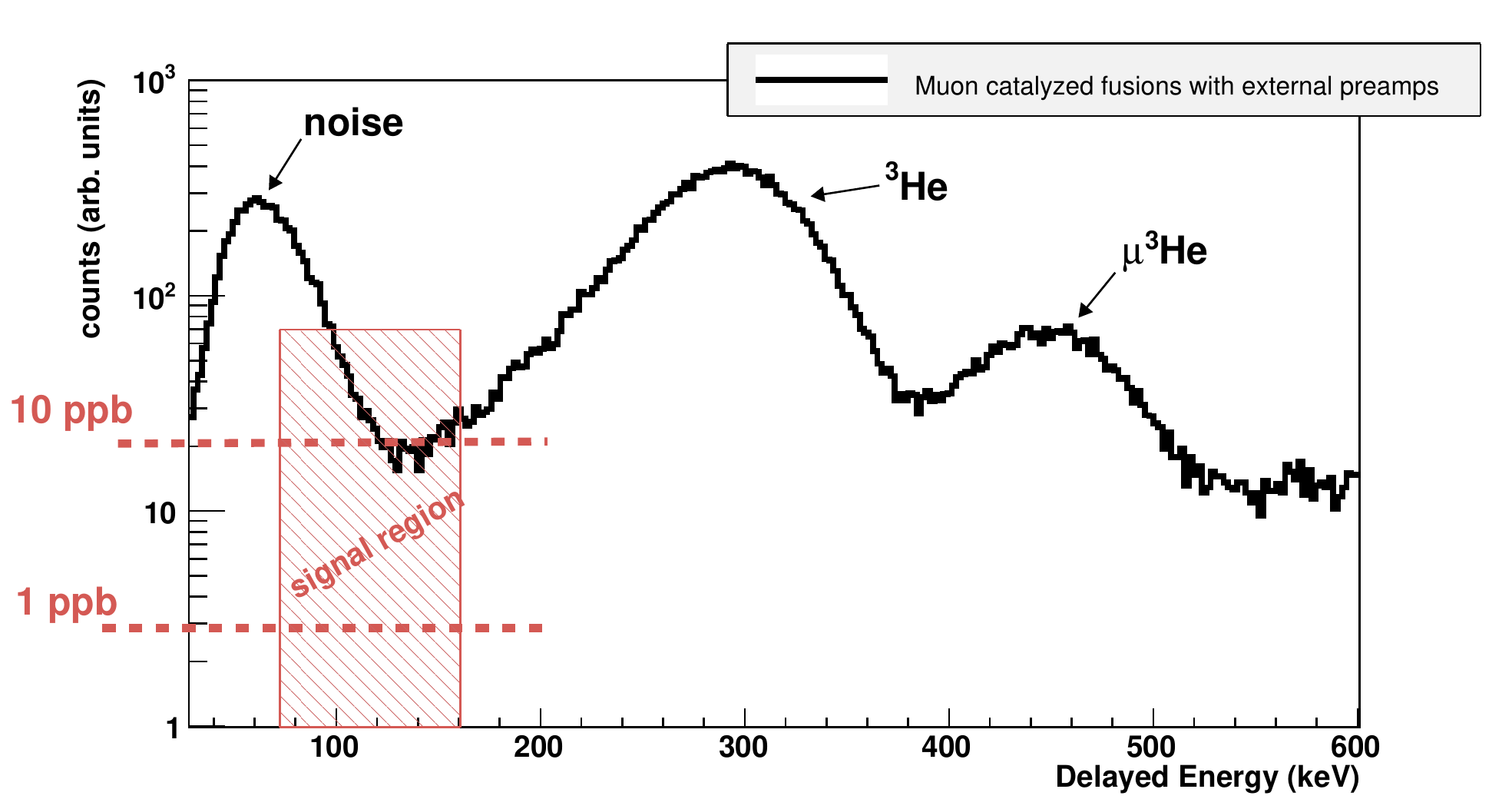}
\caption{Energy spectrum of delayed pulses following the muon stop
collected with external room-temperature preamplifiers in 2011.
The peaks are from noise, $d d \mu$ $\rightarrow$ $\mu^-$ + $^3$He + $n$, 
and $d d \mu$ $\rightarrow$ $\mu^3$He + $n$.  
The non-Gaussian low-energy tails
result from chamber effects that are described in the text.
An accompanying Michel electron is required in this spectrum, thereby 
emphasizing the muon-catalyzed fusion pulses; this is the background that 
must be subtracted in the search for nuclear recoils.
The red dashed region indicates 
the energy range where the central two-thirds of nuclear recoil events 
following muon capture on impurities would be expected.  
The horizontal lines indicate the number of events that would be expected for 
nitrogen contamination at the levels of 10 ppb and 1 ppb, so sensitivity
was limited to the 10~ppb level.
}
\label{fig:R43He}
\end{figure}

The cryogenic TPC is placed in an insulating vacuum.  In initial runs of the 
experiment, external room-temperature preamplifiers were mounted outside 
the vacuum as shown in Figure~\ref{fig:overall}.  Each pad requires a separate 
amplifier, so there are 48 channels at each stage of the electronics chain.
This arrangement required a 1~m in-vacuum cable to the nearest accessible 
flange.  In order to limit its contribution to the detector capacitance, this
long flexible Kapton ribbon cable was not shielded.
Figure~\ref{fig:R43He} shows that the sensitivity to impurities with
this external preamplifier was limited to the 10~ppb level.
In this paper, we describe our development of an in-vacuum cryogenic 
preamplifier system that allowed the long unshielded cable to be eliminated.
The new preamplifier design improved the electronic part of the resolution 
of the detector by a factor of 3, and the use of shielded cables eliminated 
interference from microphonic pickup of the drift cathode high voltage on the 
long cable.

\begin{figure}[t]
\begin{center}
\includegraphics[width=0.8\textwidth]{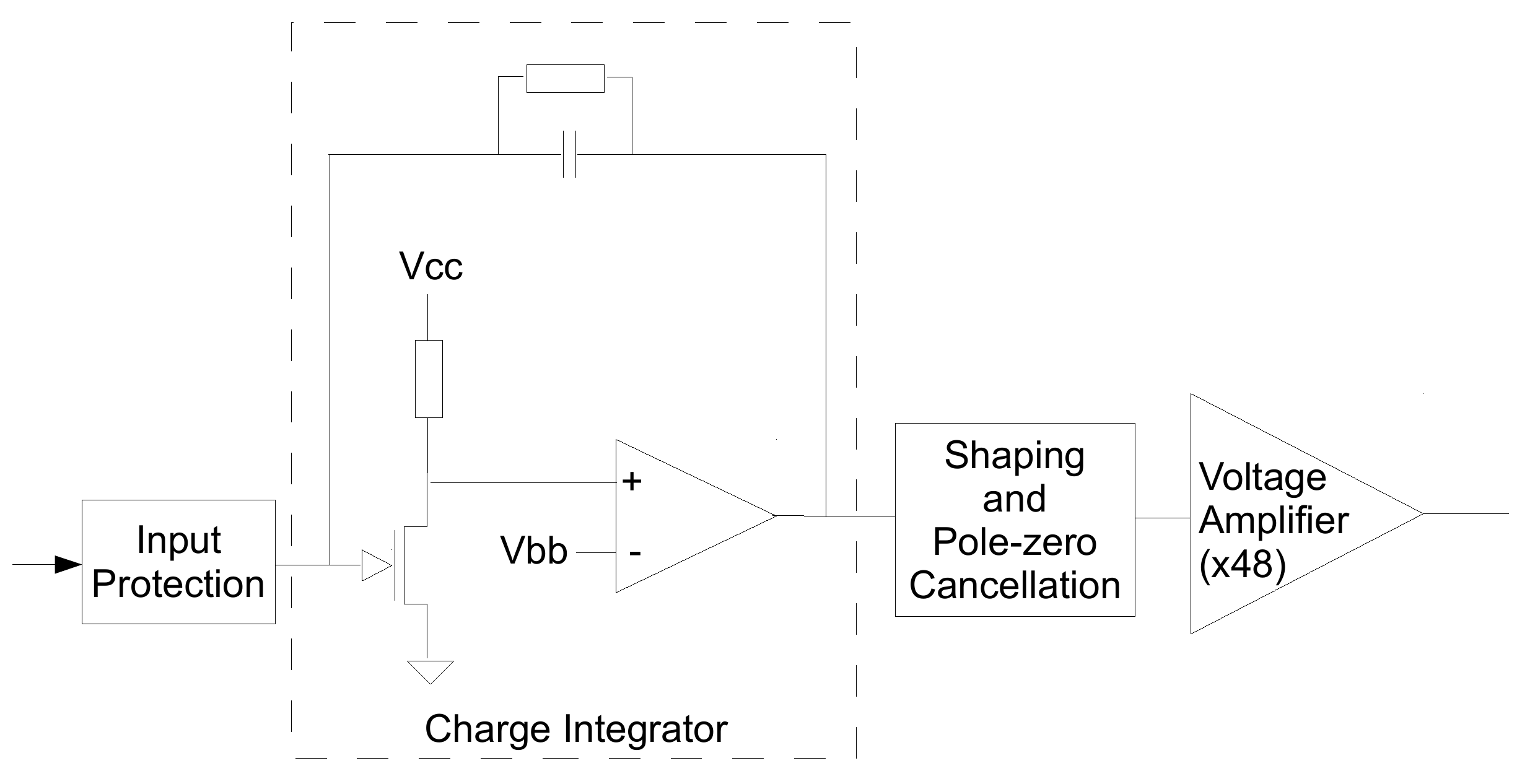}
\end{center}
\caption{Block diagram illustrating the design of the preamplifier.}
\label{fig:block}
\end{figure}

A block diagram of the new preamplifier is shown in Figure~\ref{fig:block}. 
A pair of diodes clamps the input signal voltage to protect the amplifier 
from currents induced by high voltage discharges inside the TPC.  
The first amplifying stage integrates the charge that reaches the gate of 
a junction field-effect transistor (JFET) that is inside a feedback loop 
regulated by an operational amplifier.  A first level of pulse shaping 
is required for compatibility with the existing shaping amplifiers; it
is applied before the the output is driven by a voltage gain stage.

Reference~\cite{Spieler} discusses low-noise electronics; it is placed in 
the context of semiconductor detectors, but many of the same concepts
apply here.  Drift chambers are discussed
in~\cite{Blum}, while~\cite{Hilke} reviews the current status of
time projection chamber technology.  

\section{Energy resolution}
\label{sec:resolution}

Ionization of deuterium requires $W=36.5 \pm 0.3$~eV per electron/ion pair~\cite{icru,Smirnov2005474}, 
an order of magnitude larger than in a semiconductor detector.
A 1~MeV energy deposition in deuterium therefore corresponds to $N_e =$27400 
electrons.  For such an event, the anticipated Fano factor 
of $F=0.3$~\cite{Bronic} would give as a physical limit an RMS resolution 
of $\sqrt{N_e F}$ = 90 electrons or 3.3~keV(D$_2$).  Recombination of 
electrons and ions due to space-charge density effects is minimal for muon 
tracks.  Approximately half of the charge deposited by a nucleus such 
as ${}^3$He is lost to this recombination, because of the large $dE/dx$.  
Recombination at that level would increase the optimal 
physical resolution to 110 electrons or 4.0~keV(D$_2$) RMS for that peak.  
However, it is widened much further by a long non-Gaussian tail on the 
low-energy side, because the recombination fraction depends on 
the angle of the ionization track relative to the applied electric 
field~\cite{argoneut}.
It was therefore determined that a reasonable design goal would be a 
resolution at the level of 10~keV(D$_2$) RMS, which would provide a   
substantial improvement relative to the 29~keV(D$_2$) RMS
of the existing room-temperature preamplifier.

\begin{figure}
\includegraphics[width=\textwidth]{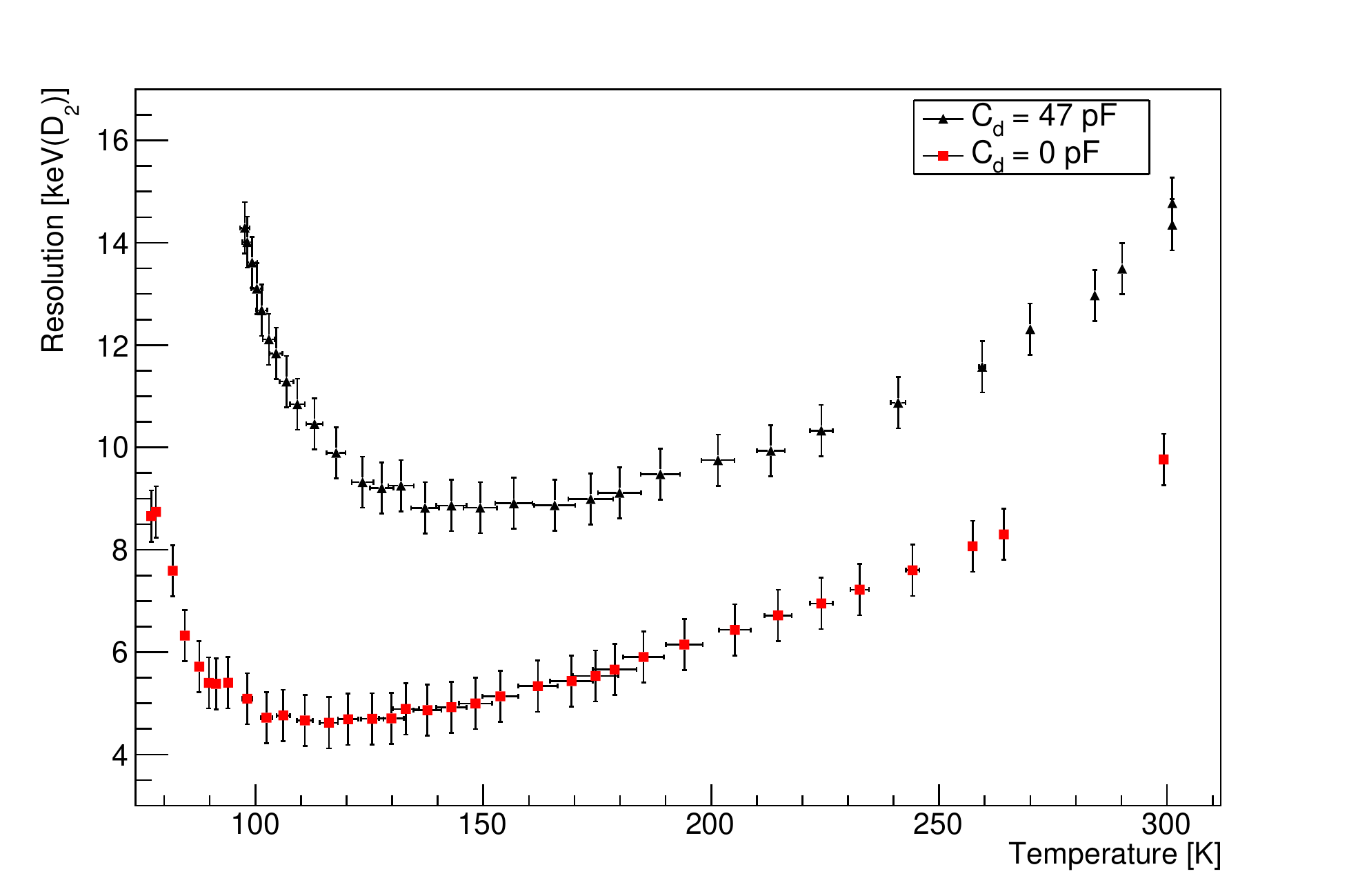}
\caption{Temperature dependence of energy resolution of prototype board 
at 1.1~{\textmu}s external integration time, with and without simulated detector 
capacitance, without input protection.}
\label{fig:tempScan}
\end{figure}

Tests of the amplifier were performed by two groups at different 
institutions with somewhat different laboratory setups.  The best purely 
electronic resolution that was observed for the new amplifier 
was 4.5~keV(D$_2$), or 120 electrons, RMS.  This result was obtained 
for a prototype where the protection diodes were omitted and no detector 
was connected.  In this case, the preamplifier was cooled slowly by 
suspending it in the cold vapor just above the liquid level in a nitrogen 
dewar.  An Ortec 485 amplifier with a shaping time of 1.1~{\textmu}s (rather 
than the custom shaping amplifiers from the experiment) and a multichannel 
analyzer were used to collect the charge spectrum.  A square-wave pulse 
was injected through the test input (a resistor divider and capacitive 
coupling to each channel), 
simulating approximately 1~MeV(D$_2$) of energy deposition.  
The amplifier reached the optimal resolution at a temperature near 115~K.
As shown in Figure~\ref{fig:tempScan}, the resolution for the same
amplifier at room temperature was 10~keV(D$_2$) RMS.
When a 47~pF ceramic capacitor was added, somewhat larger than the actual
pad plane and cable capacitance of $\sim$10~pF, the optimal temperature 
increased slightly to 140~K.
At this point, the resolution was 10~keV(D$_2$) RMS; it was 16~keV(D$_2$) RMS at 
room temperature.  Comparable results were obtained by the other 
group.

The MuSun electronics must be optimized for a short shaping time, which is 
needed to resolve delayed fusion and impurity capture pulses from the initial 
muon stop pulse.  These delayed pulses begin
arriving immediately and decay with a fast component of $\sim$0.4~{\textmu}s  
and a slow component of the 2.2~{\textmu}s muon lifetime.  
The output of the preamplifier therefore drives a shaping amplifier that 
is a fifth-order Bessel filter with a 0.5~{\textmu}s integration time; this device 
also includes a baseline restorer to remove low-frequency microphonic pickup. 

\begin{figure}
\includegraphics[width=\textwidth]{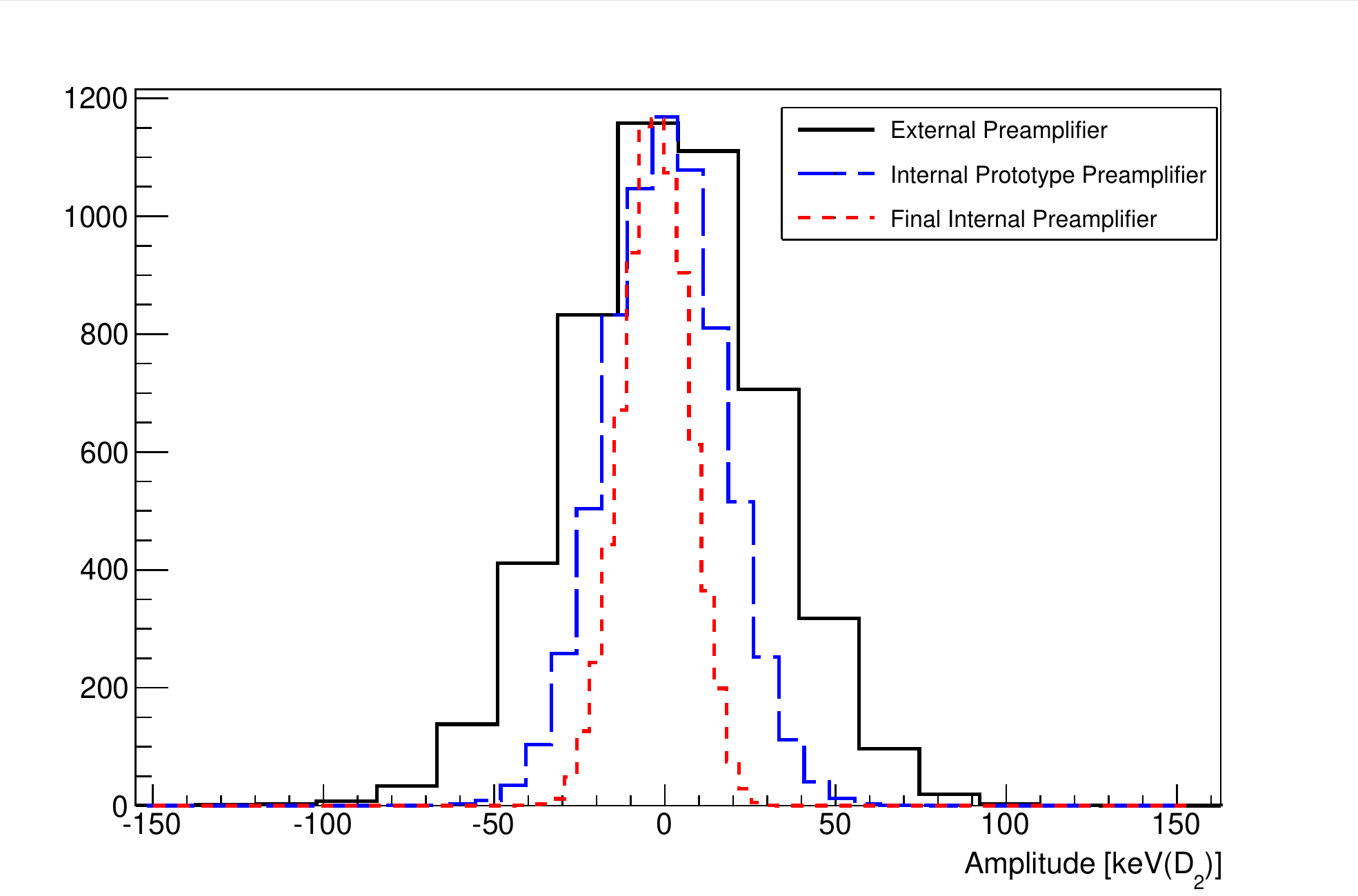}
\caption{Comparison of pulse amplitude distributions (in units of keV, 
centered at zero) for a test pulse with an external room-temperature 
preamplifier, with prototype cryogenic amplifier at 260~K, and 
with the final cryogenic amplifier at 140~K as installed in the experiment.}
\label{fig:amp_distro}
\end{figure}

\begin{figure}
\includegraphics[width=\textwidth]{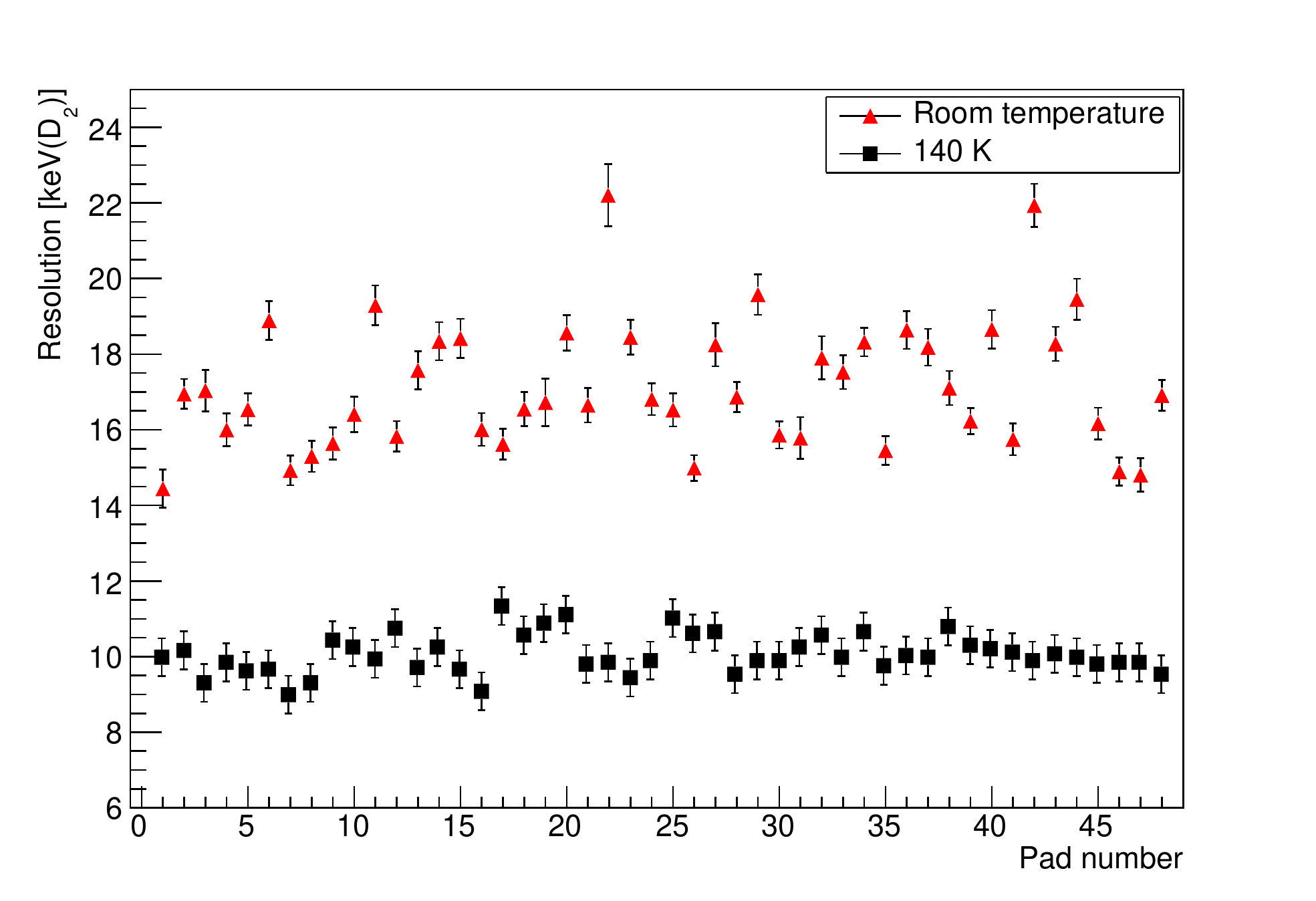}
\caption{Measurements of the preamplifier resolution for all 48 pads
in situ at room temperature (red triangles) and at the operating temperature of
140~K (black squares).}
\label{fig:in_situ}
\end{figure}

In the experimental environment, the outputs of the shaping amplifier are
connected to waveform digitizers that sample the signal at 50~MHz, and the 
pulse waveforms are fit to a template to extract the amplitude and integral.
When the signal source is an electronic test pulser, the 
resulting spectrum is a Gaussian, illustrated in Figure~\ref{fig:amp_distro}.
The result of fitting this distribution for each of the 48 pads 
is shown in Figure~\ref{fig:in_situ}.  The results reproduce those seen in 
the prototype laboratory tests: at room temperature, the typical resolution 
is 16~keV(D$_2$) RMS, which improves to 10~keV(D$_2$) RMS at the operating 
temperature of 140~K.  

\section{Amplifier design} 

\begin{figure}
\begin{center}
\includegraphics[width=0.9\textheight,angle=90]{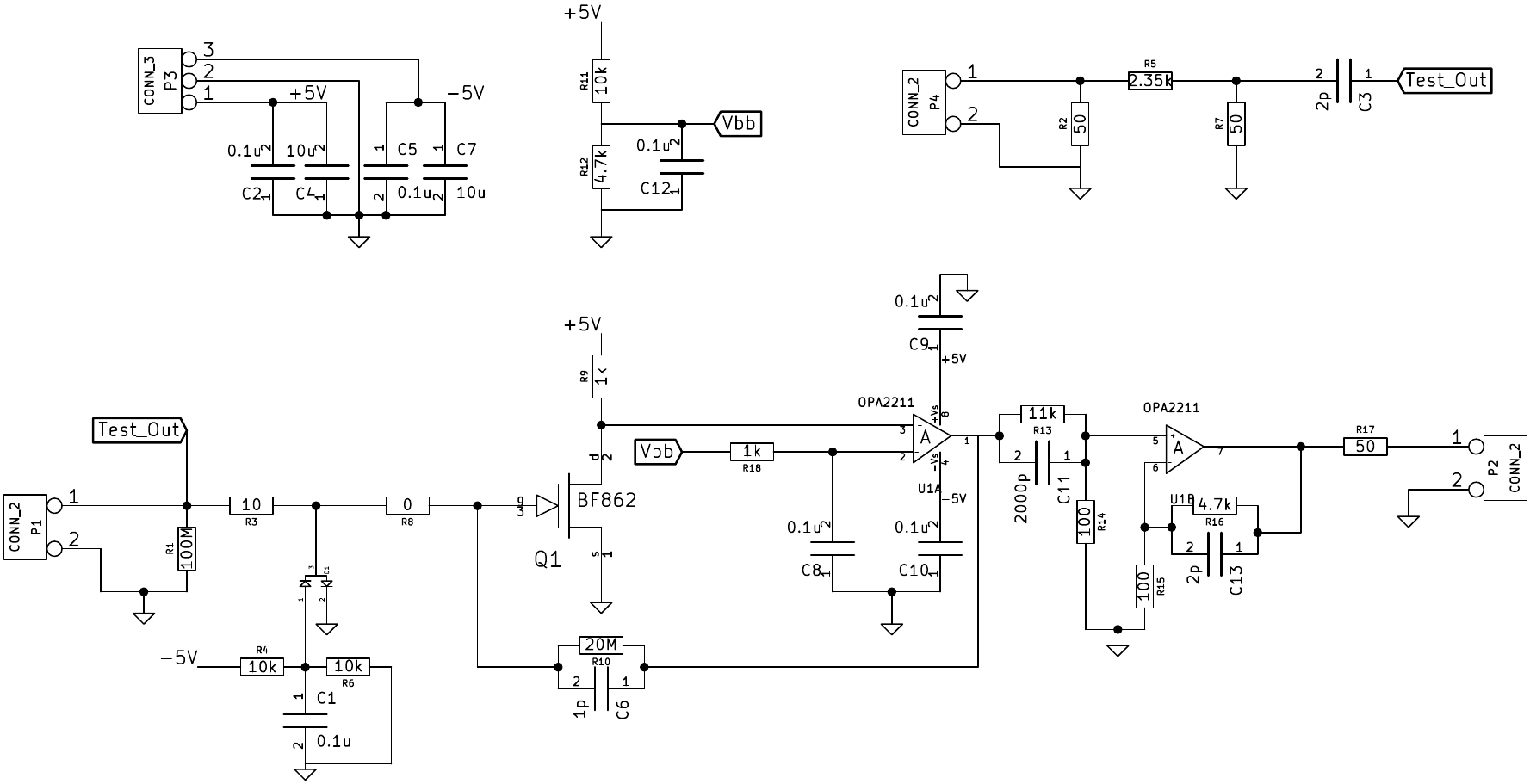}
\end{center}
\caption{Full schematic of one channel of the preamplifier.}
\label{fig:schematic}
\end{figure}

The full schematic of the new amplifier is shown in Figure~\ref{fig:schematic}.
A pair of BAV199 (NXP Semiconductors) diodes provides input protection.
The first amplifying stage integrates the charge that reaches the gate of 
a BF862 (NXP Semiconductors) JFET, which is in a common-source configuration.
It is inside a feedback loop regulated by one channel of an OPA2211 (Texas 
Instruments) operational amplifier, which maintains the quiescent bias
of the JFET:
$V_{gs} = V_{bb} = V_{cc} \cdot R_{11}/(R_{11} + R_{12}) = 1.6~{\rm V}$, 
$I_{ds} = (V_{cc} - V_{gs})/R_9 = 3.4~{\rm mA}$.
The pulse is shaped by a first stage 
of $CR$ filtering with pole-zero cancellation.  Finally, the output is driven 
by a non-inverting voltage amplifier with a gain of 48.

\subsection{Input protection}

The input protection circuit has allowed the amplifiers to work reliably in an 
environment where discharges occur sporadically from the drift cathode to 
grounded components, inducing voltages on the pads. 
Initial prototypes used the BAV99 protection diode (a common part number; 
devices from Fairchild Semiconductor and NXP Semiconductor exhibited 
similar behavior), which increased the noise at room temperature.
The BAV199 that was eventually employed is specified for three orders of 
magnitude lower leakage current and therefore makes a negligible 
contribution to the noise.

\subsection{Field effect transistor}

\begin{figure}
\begin{center}
\includegraphics[width=0.8\textwidth]{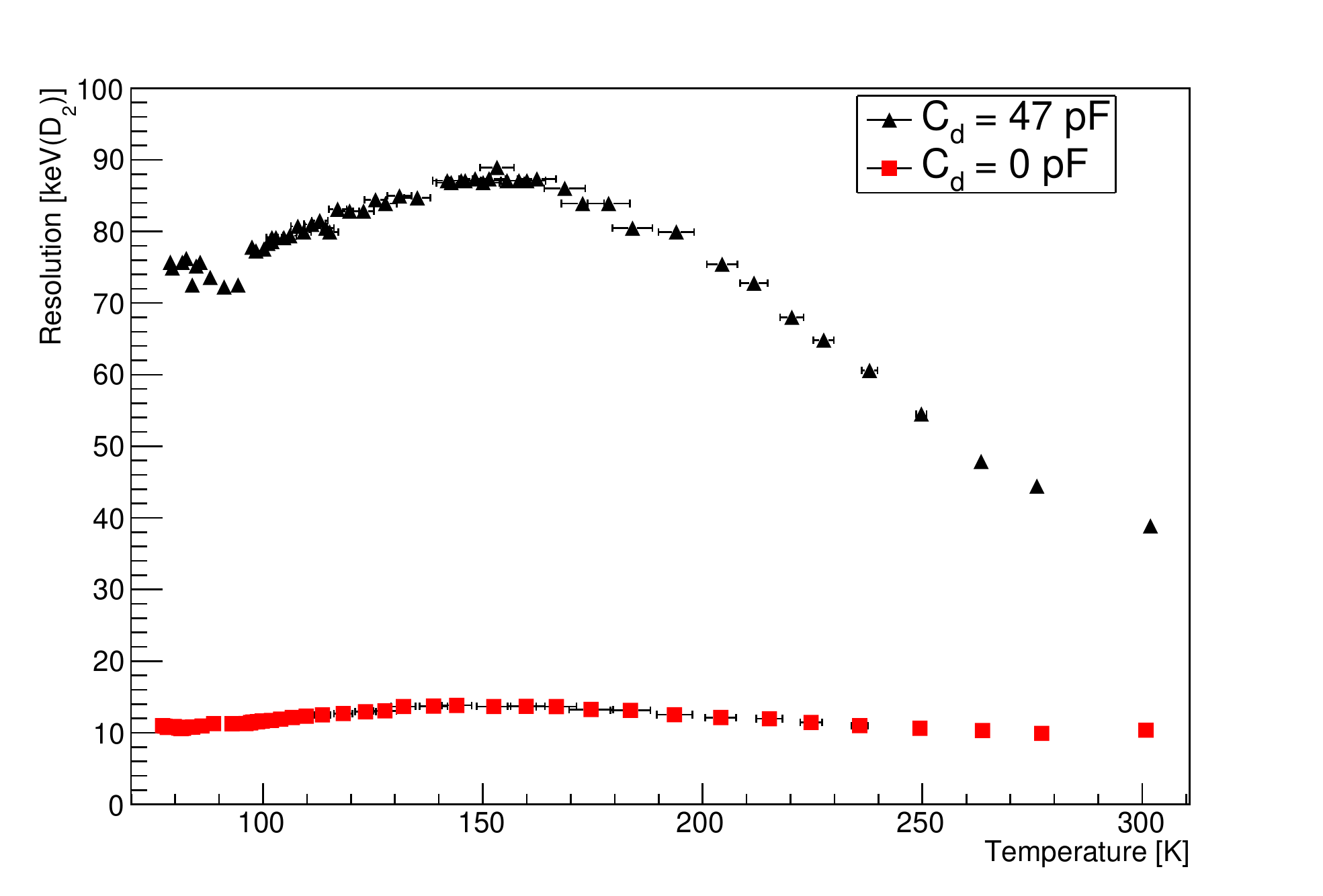}
\end{center}
\caption{Energy resolution versus temperature for BF998 dual-gate MOSFET
(a device that was not selected), with and without a simulated detector 
capacitance.}
\label{fig:bf998}
\end{figure}

The most critical component that determines the noise performance is the 
field-effect transistor on the front end.  Consequently, the design of the 
amplifier required testing and characterization of potential transistors.
Other models that were considered included the BF998 dual-gate MOSFET and 
the BF861A and BF861B JFETs (NXP Semiconductor).  
Performance of the design with all of these devices was significantly 
worse than with the BF862 that was eventually chosen.  As an example to 
show the scale of the effect, the resolution for the BF998 is shown 
in Figure~\ref{fig:bf998}.   While this device seemed promising at
room temperature with no input connected, the noise slope versus 
detector capacitance was not acceptable, and it did not improve 
with cooling.

\subsection{Operational amplifier}

The OPA2211 operational amplifier component was selected because we had 
successfully tested its single-channel counterpart, the OPA211, cooling
it down to a temperature of 65~K where it turns off.  Also, it is specified 
for a low voltage noise of 1.1~nV/$\sqrt{\rm Hz}$ at room temperature.
Excellent cryogenic performance was expected for it based on the BiCom3HV 
silicon-germanium (SiGe) heterojunction process~\cite{BiCom3HV} used in its 
fabrication.  Individual SiGe transistors have been used in applications 
requiring temperatures as low as 4~K~\cite{SiGe}.  

However, prototypes of two alternate designs of the amplifier were constructed 
and evaluated for comparison.  One used the Amptek A250 low-noise amplifier 
block in place of the OPA2211 in the charge-integrating stage, and one replaced 
it with an amplifier with three discrete bipolar junction transistors. 
The A250 amplifier was tested at room temperature, and the transistor 
amplifier was tested both at room temperature and under cryogenic conditions.  
Both of these alternate designs provided energy resolutions that were 
essentially the same as the one that was eventually chosen.  However, 
the OPA2211 is a much less expensive component than the A250, and the discrete 
transistor amplifier has a larger part count, which makes the circuit board 
layout less compact and may reduce its reliability.


\subsection{Passive components}

Care was used in the selection of resistors and capacitors.  Thin-film 
and MELF resistors were used in preference to thick-film, which are
known to exhibit excess noise above the Johnson-Nyquist limit.   
NP0/C0G dielectric, which has minimal temperature dependence, was preferred 
for all capacitors except some of the larger power filter bypass capacitors.

\subsection{Shaping time constants}

The dominant vibration frequency of the apparatus, which was driven 
by the cold-head compressor, is approximately 500~Hz; other modes 
with frequencies in the kHz have been seen.   With the external preamplifiers,
this frequency was picked up by the Kapton cable because it traveled parallel 
to the -80~kV drift cathode high voltage, creating a capacitor whose 
capacitance varied with vibration.  However, similar pickup may originate 
inside the TPC between the Frisch grid and the pad plane.  This acoustic 
frequency must be rejected at an early stage so that the preamplifier is not 
saturated by the oscillating baseline, which was previously seen at the level 
of several volts at the output.

The differentiation time constant between the two stages of the preamplifier
is set to provide this filtering.  It is determined by 
$t_{shape} = R_{14} \cdot C_{11}$ = 0.2~{\textmu}s, so the -3~dB point
of the highpass filter is at $f_{shape} = 1/2\pi t_{shape}$ = 800~kHz.  
This provides substantial separation from the acoustic frequencies 
so that they are suppressed by many orders of magnitude.

A requirement of quick recovery to prepare for the next pulse,
along with additional filtering of the acoustic frequencies,
motivated the choice of the feedback resistor $R_{10}$=20~M$\Omega$.  
While a larger feedback resistance would in principle introduce less noise,
the discharge time constant of $R_{10} \cdot C_{6} = 20$~{\textmu}s is matched to 
the requirement that the dynamic range of the charge-integrating
stage  should not be saturated when the next muon enters the TPC, 
typically 25~{\textmu}s later.  The pole-zero time constant 
$R_{13} \cdot C_{11}$=22~{\textmu}s was tuned experimentally to optimize the 
cancellation of the long tail of the output pulse.

\section{Noise spectrum and shaping time}

\begin{figure}
\begin{center}
\includegraphics[width=0.8\textwidth]{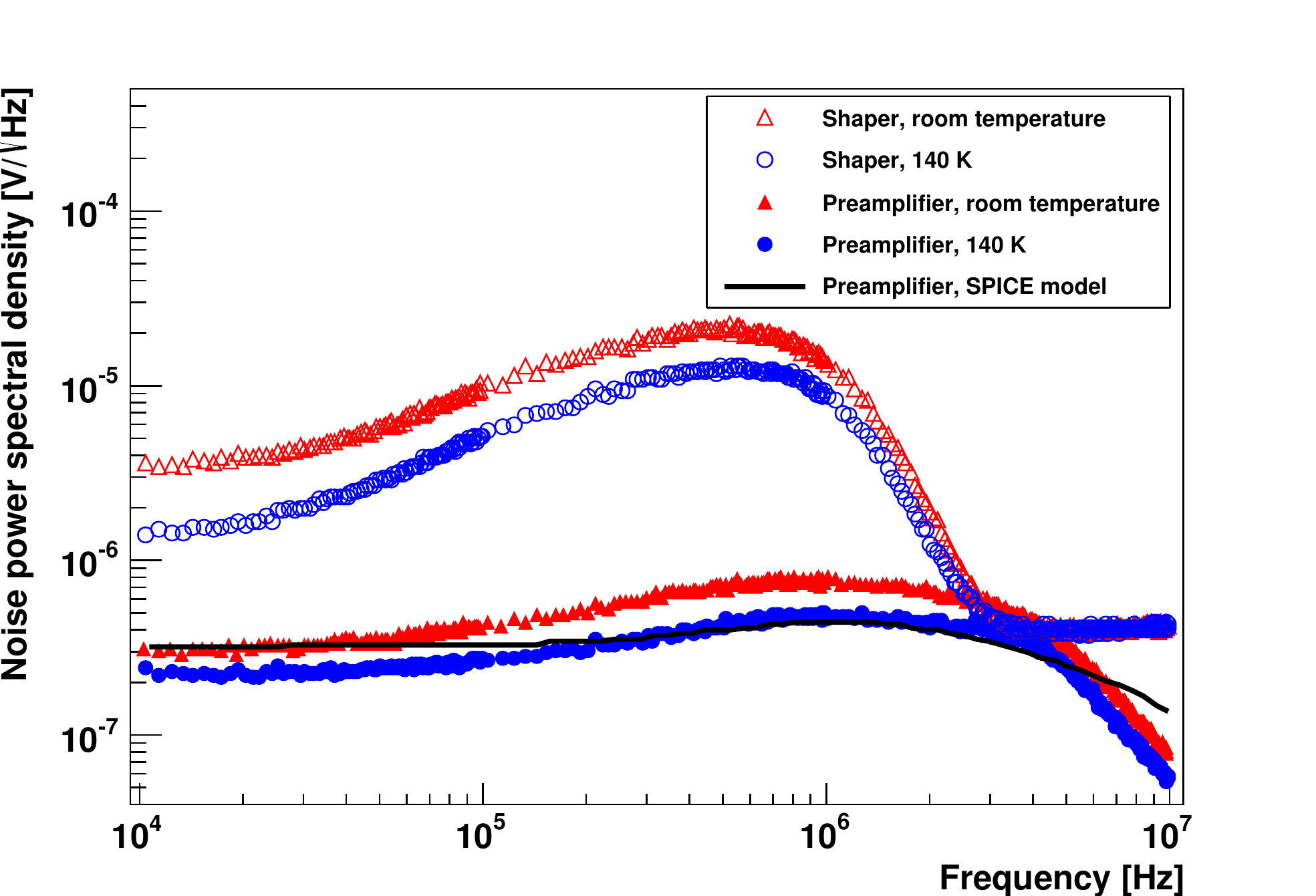}
\end{center}
\caption{Noise power spectral density measured from Fourier transforms of 
signals recorded with a digital oscilloscope.  The spectrum is shown for warm 
and cold conditions of the preamplifier, at the output of the preamplifier and after
the shaping amplifier.  An LTspice~\cite{LTspice} circuit simulation
model prediction of the noise at the preamplifier output is also shown.}
\label{fig:noise_psd}
\end{figure}

In order to interpret the noise of the preamplifier in terms of
its frequency components, the noise power spectral density (PSD) was 
measured from Fourier transforms
of the autocovariance of waveforms recorded with a digital oscilloscope.
This noise spectrum, shown in Figure~\ref{fig:noise_psd}, agrees 
with a LTspice~\cite{LTspice} circuit simulation model in the general 
shape, and it quantitatively agrees within a factor of $\sim$2.
The model indicates that the largest contribution to the low-frequency 
voltage noise is from the parallel resistance $R_{10}$, which discharges
the charge-integrating stage.  The largest contribution to the high-frequency
side is the input voltage noise of the JFET.

We used the PSD 
to predict the dependence of the resolution on the external shaping time.
For this purpose, the PSD was measured with $C_{11}$ bypassed to remove the 
early differentiation stage.
The measured spectrum $e_n(\omega)$ was folded with the transfer 
function $A_v(\omega)$ of a $CR-RC^n$ shaping amplifier and integrated 
over frequency~\cite{Spieler}
\begin{eqnarray}
v_{n}^2 &=& \frac{1}{2\pi} \int\limits_{-\infty}^{\infty} e^2_{n}(\omega) ~ | A_v(\omega) |^2 d\omega ~, \\
A_v(\omega) &=& \frac{j \omega \tau}{(1 + j \omega \tau)^{n+1}} ~, 
\end{eqnarray}
to give a prediction of the total effective voltage noise $v_n$ as a 
function of the shaping time $\tau$.  
In this calculation, $n$ was taken to be 3, which matched the observed
output pulse shape from the shaping amplifier.

\begin{figure}
\begin{center}
\includegraphics[width=0.8\textwidth]{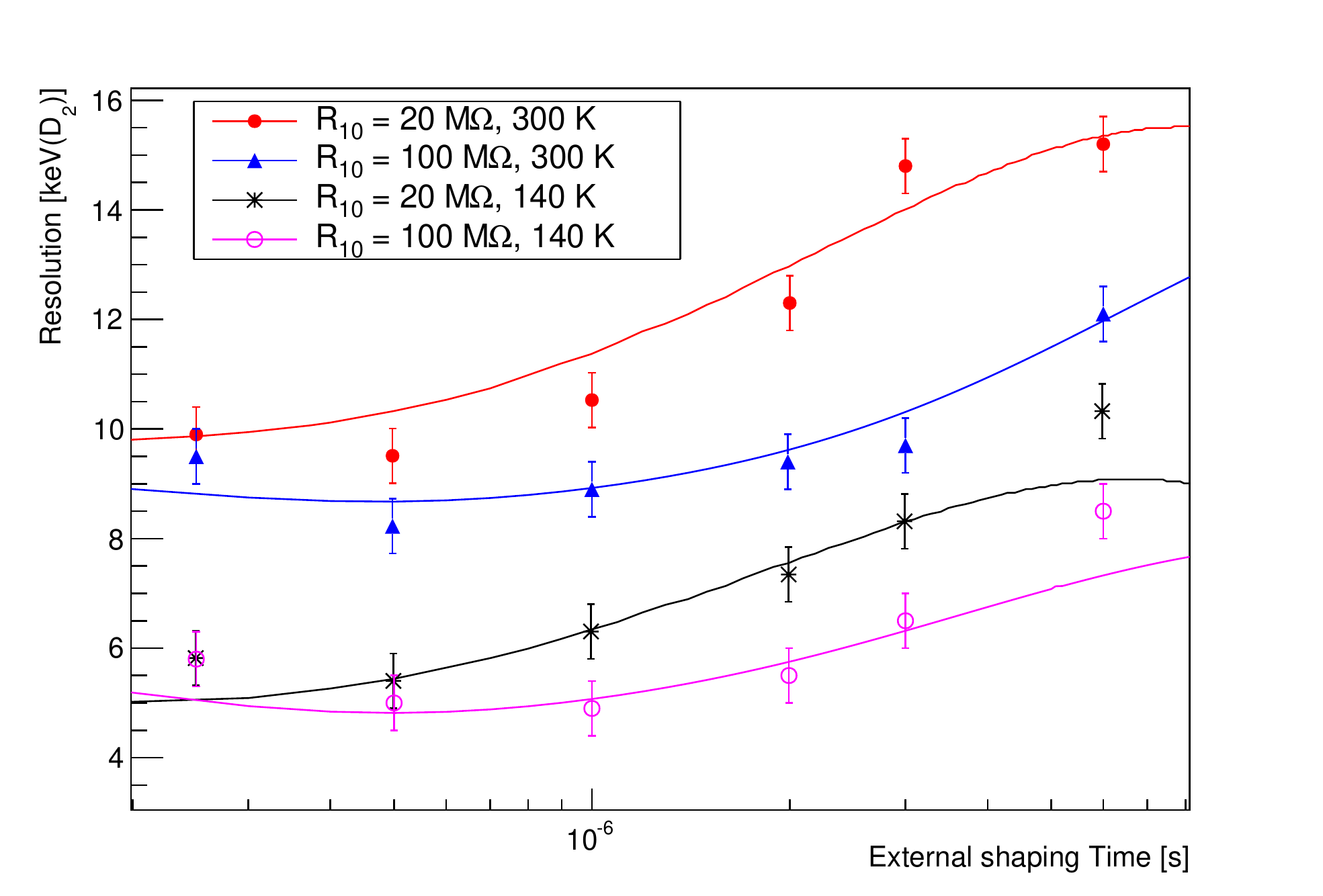}
\end{center}
\caption{The resolution calculated from the noise power spectral density 
(smooth curves) is compared to direct measurements (discrete data points)
as a function of shaping time for preamplifier channels, with 20~M$\Omega$ 
and 100~M$\Omega$ discharge resistors and at both room temperature and 140~K.
No detector capacitance was connected.}
\label{fig:enc_shaping}
\end{figure}

After normalization to units of energy, $v_n$
appears in Figure~\ref{fig:enc_shaping}.  This integration
demonstrates that the noise spectrum corresponds to an optimal 
shaping time near the 0.5~{\textmu}s that is required, which was confirmed
by direct measurements of the resolution as a function of the 
shaping time set on an Ortec 673 amplifier.
These measurements were performed with both 20~M$\Omega$ and 100~M$\Omega$
discharge resistor ($R_{10}$) values.  They show that the impact on 
the resolution of the 20~M$\Omega$ discharge resistor is minimal for 
short shaping times.

\begin{figure}
\begin{center}
(a)
\includegraphics[width=0.9\textwidth]{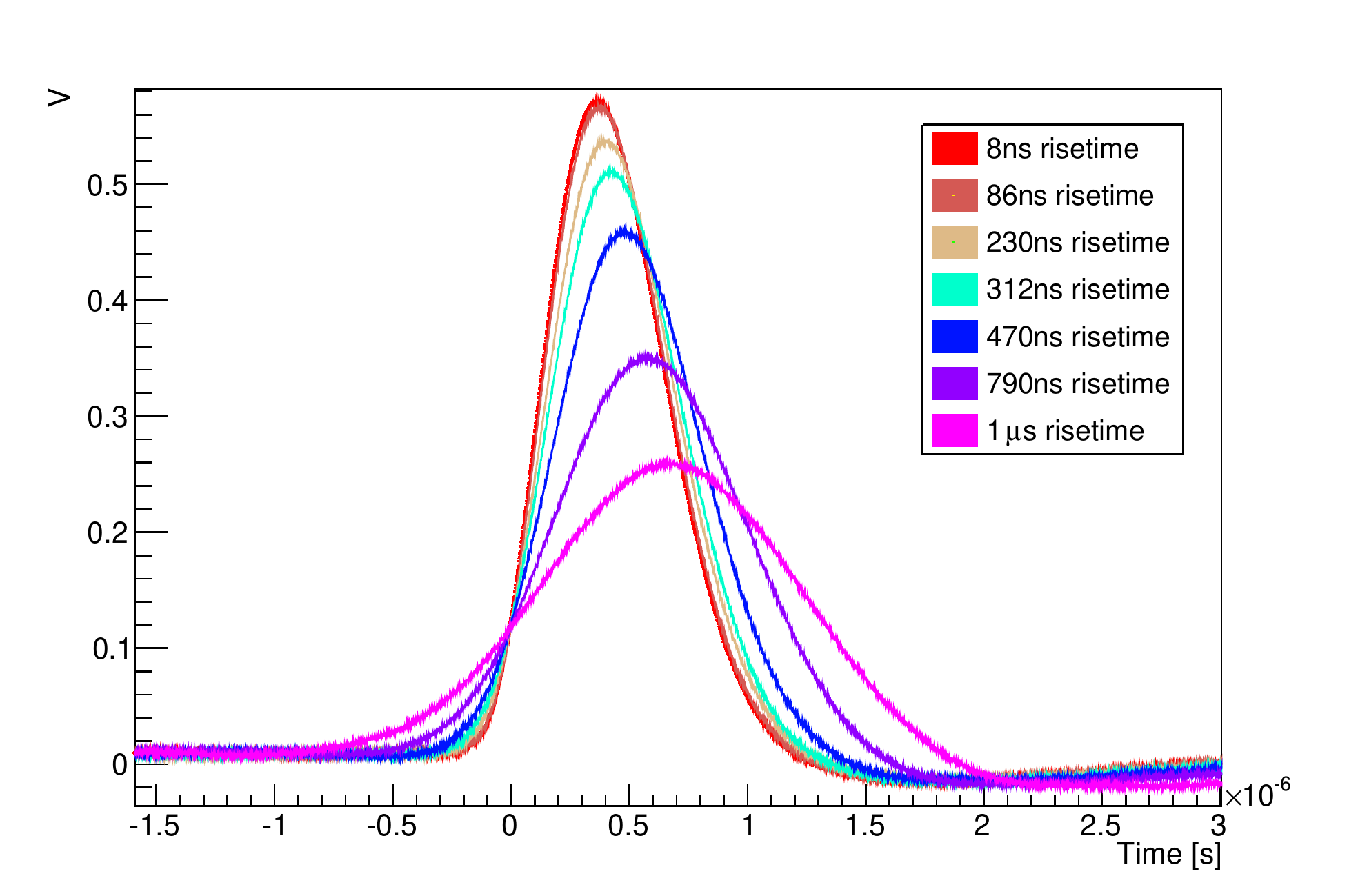}

(b)
\includegraphics[width=0.8\textwidth]{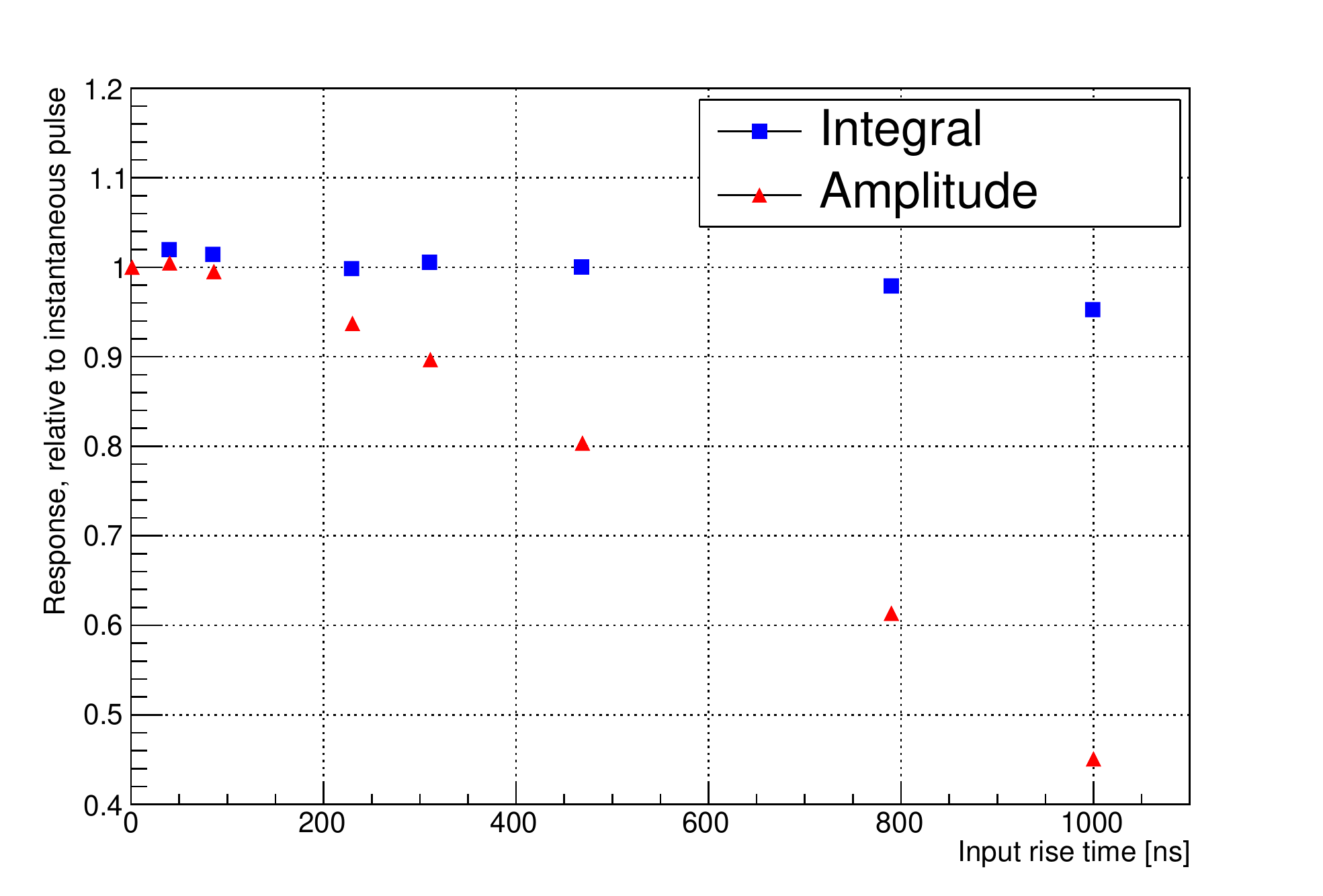}
\end{center}
\caption{(a) Pulse shapes from the output of the shaping amplifier for varying
charge rise times, as recorded by a digital oscilloscope.
(b) Dependence of the amplitude and integral of the pulse on the 
charge rise time, measured relative to the shortest rise time.}
\label{fig:risetimes}
\end{figure}

The short shaping time leads to a dependence of the width of the 
shaped pulse on the rise time over which the input charge is collected,
as shown in Figure~\ref{fig:risetimes}.  This leads to a significant 
reduction in the amplitude: 55\% over 1~{\textmu}s. 
However, because the full waveform of the output of the shaping amplifier 
is recorded, it is possible in the analysis to integrate the pulse shape.
The integral remains stable versus risetime for 0.5~{\textmu}s and 
decreases by only 5\% over 1~{\textmu}s.

\section{Installation and reliability}

Four channels of a prototype version of the amplifier were deployed
in an engineering run in 2012.  These prototypes had only two channels
per board, so the layout was not as compact as the final version, and some 
component values were different.  The shaping time constant
on the preamplifier was set to 100~{\textmu}s rather than 0.2~{\textmu}s, 
effectively bypassing the internal shaping stage, in order to be able to 
use off-the-shelf amplifiers where the shaping time could be varied.
Consequently, the acoustic pickup was not fully suppressed.
They were cooled to 260~K by contact 
with the long stainless steel support rods for the TPC, which is held at 31~K.
Two channels were connected to the TPC by coaxial cables and two by 
unshielded wires.  However, even the coaxial cable was not fully shielded,
with $\sim$1~cm of bare conductor near the feedthrough to the TPC.  
Microphonic pickup was still evident; the signal amplitudes at 
the preamplifier output were $\sim$0.9~V on the coaxial channels 
and $\sim$1.8~V on unshielded channels.

\begin{figure}
\begin{center}
(a) \includegraphics[width=0.90\textwidth]{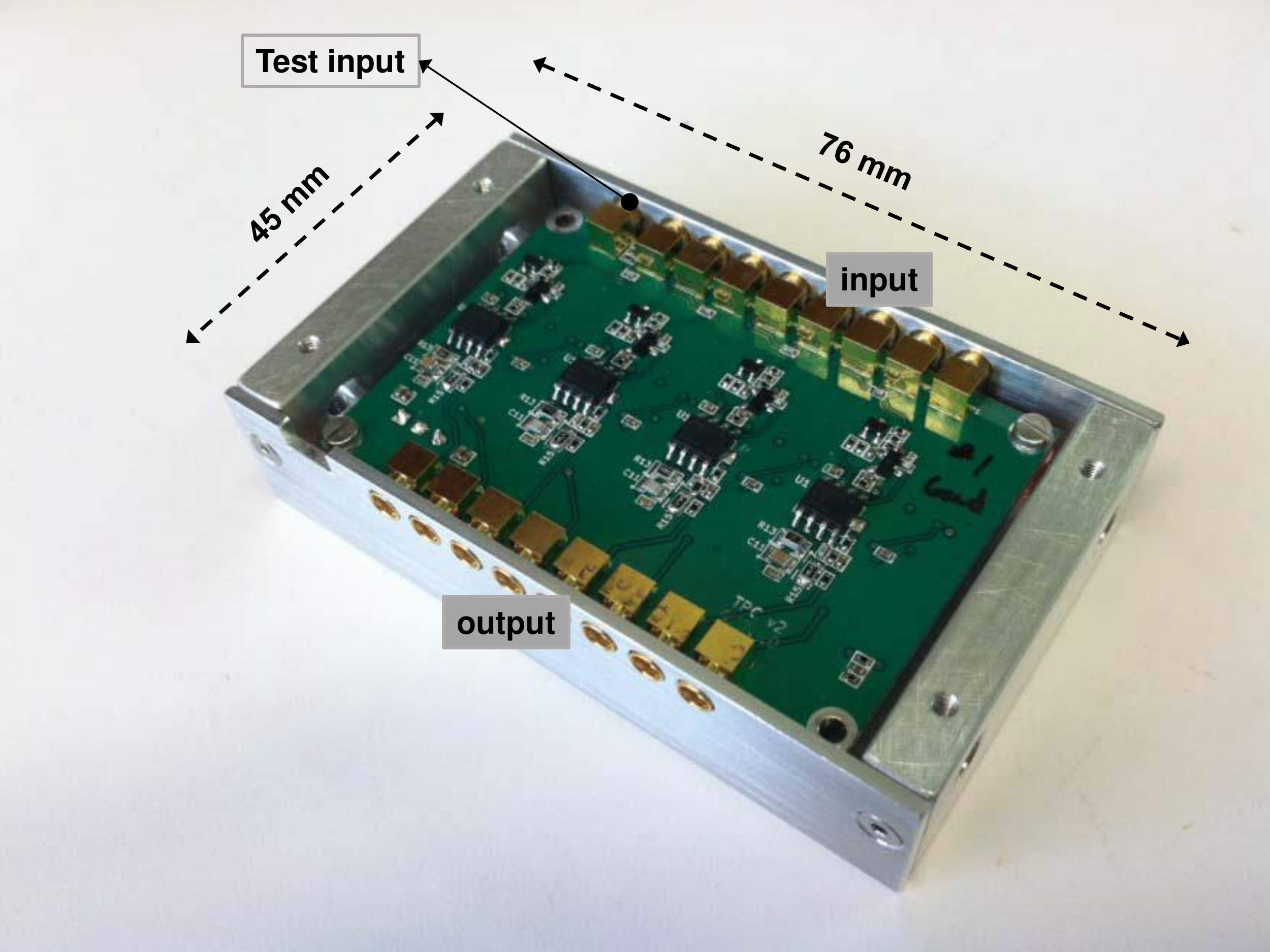}

(b) \includegraphics[width=0.90\textwidth]{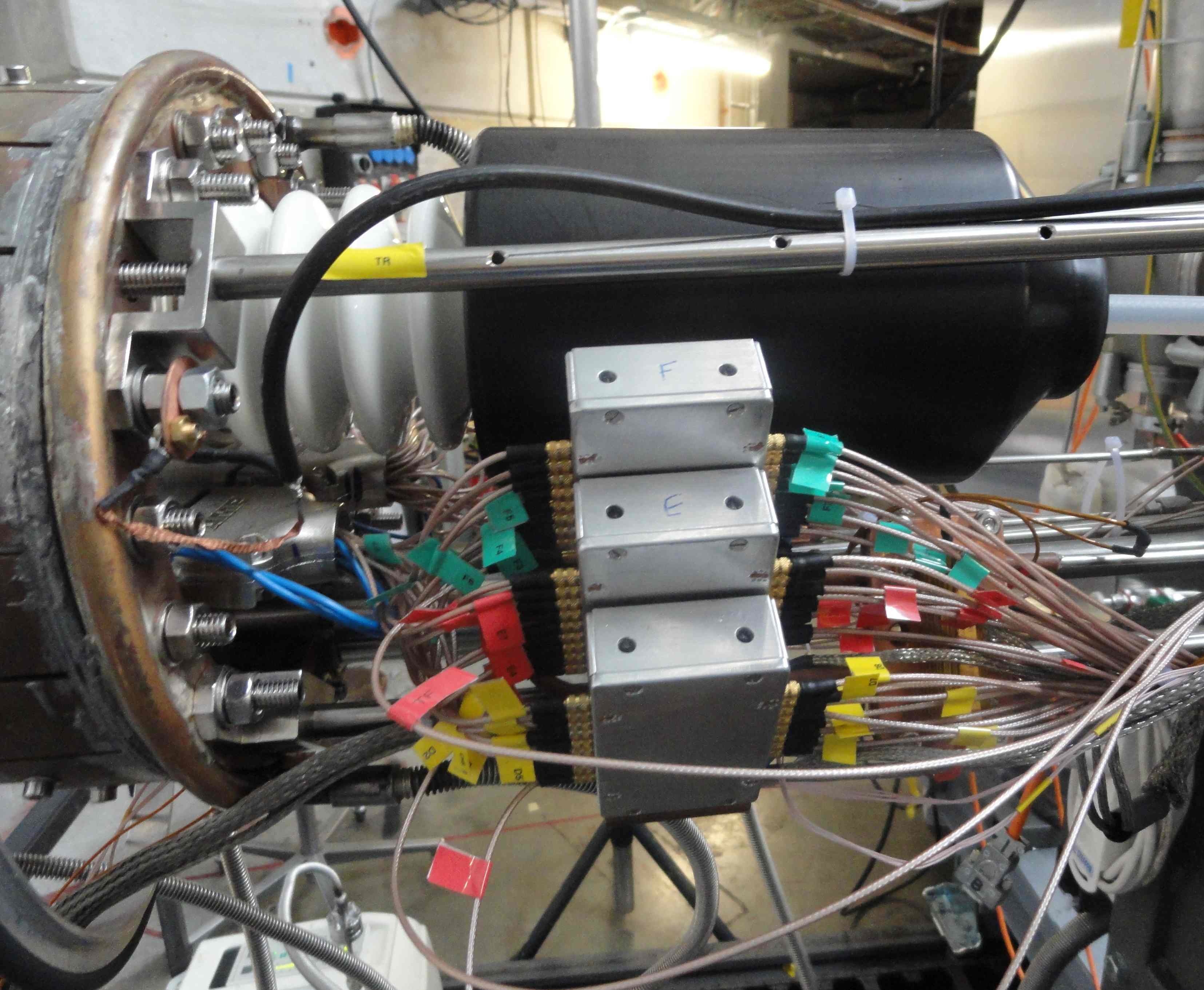}
\end{center}
\caption{(a) Photograph of one of the preamplifier boards, mounted
in an aluminum enclosure box.  (b)~Photograph of half of the preamplifiers
connected to the TPC.}
\label{fig:boardPhoto}
\end{figure}

All 48 pads were instrumented for the 2013 data collection run.
The final version of the preamplifier card had to be quite compact
to allow this number of channels to fit in the small space that
was available.  Each card holds eight channels, 
which are independent except for common power supply filtering and a 
test input.  Each board is enclosed in a grounded aluminum box,
$45 \times 76 \times 15$~mm$^3$,
as shown in Figure~\ref{fig:boardPhoto}.
The input and output signals pass through MMCX connectors on the board; 
RG178 coaxial cable is used to connect through the vacuum to DB-50 feedthrough 
connectors on the TPC and on the outside.  The signal is fully enclosed 
by a ground shield all the way to the feedthrough into the TPC; this
change, together with the change to the shaping time constant, suppressed 
the microphonic pickup to an undetectable level. 

The preamplifier boxes are 
cooled by contact with a copper block, into which cold gas from liquid 
nitrogen boiloff is introduced.  The flow of nitrogen is regulated by a 
needle valve to select the desired temperature, which is monitored by 
PT100 sensors attached to two of the boards.  The power consumption of 
the entire set of 48 channels is 3.4~W at room temperature, decreasing by a 
factor of $\sim$2 at cryogenic temperatures, so the heat is easily removed 
in this way.  
Indeed, low power consumption was one of the initial 
design requirements because we had initially envisioned cooling them
with our primary cryogenic system by thermal coupling to the 31~K deuterium, 
and only a few watts of spare capacity are available in that system.
Figure~\ref{fig:amp_distro} shows the improvement in energy resolution that
was achieved with respect to the original external room-temperature 
amplifier system and to the partially-cooled prototype.

During the main production period of the 2013 run, the preamplifiers
worked reliably; all 48 channels operated for a 48 day period.
However, previous experience demonstrated two failure modes: 
discharges from the Frisch grid to the pad, and mechanical failure
from overly rapid cooling.  To avoid these problems in the future, the 
Frisch grid will be rebuilt with higher precision, and the protocol 
for cooling limits the rate to $\sim1$~K per minute.

\clearpage

\section{Conclusion}

The performance of the new MuSun preamplifier compares well to
other preamplifiers used in state-of-the-art nuclear physics experiments.  
The resolution was measured to be 4.5 keV(D$_2$) (120 electrons) RMS with 
no detector capacitance and a 1.1~{\textmu}s shaping time, 
or 10 keV(D$_2$) (250 electrons) RMS with a 0.5~{\textmu}s shaping time in the 
experimental environment.  As a point of comparison, the preamplifier designed 
for the GERDA neutrinoless double beta decay experiment reported 3~keV(Ge) FWHM
resolution with a 33~pF detector capacitance at a 1~{\textmu}s shaping time 
in liquid argon~\cite{GERDA-amp} at 87~K.  This resolution corresponds 
to 430 electrons RMS or 16~keV(D$_2$); it improves by a factor of $\sim$3 
at the longer shaping times for which it is optimized.  The commercial 
thermoelectrically cooled Amptek A250CF is advertised with a resolution 
of ``$\sim$76 electrons RMS,'' or 2.8~keV(D$_2$), but at a shaping time 
of 2~{\textmu}s~\cite{Amptek}.

Reliable operation of the preamplifiers is essential, since they 
are time-consuming to access for repair.  This reliability was achieved 
in the 2013 MuSun data collection period.  The full set of 48 preamplifier 
channels operated reliably for a continuous period of 48 days of data 
collection and technical and systematic studies.

\begin{figure}
\includegraphics[width=\textwidth]{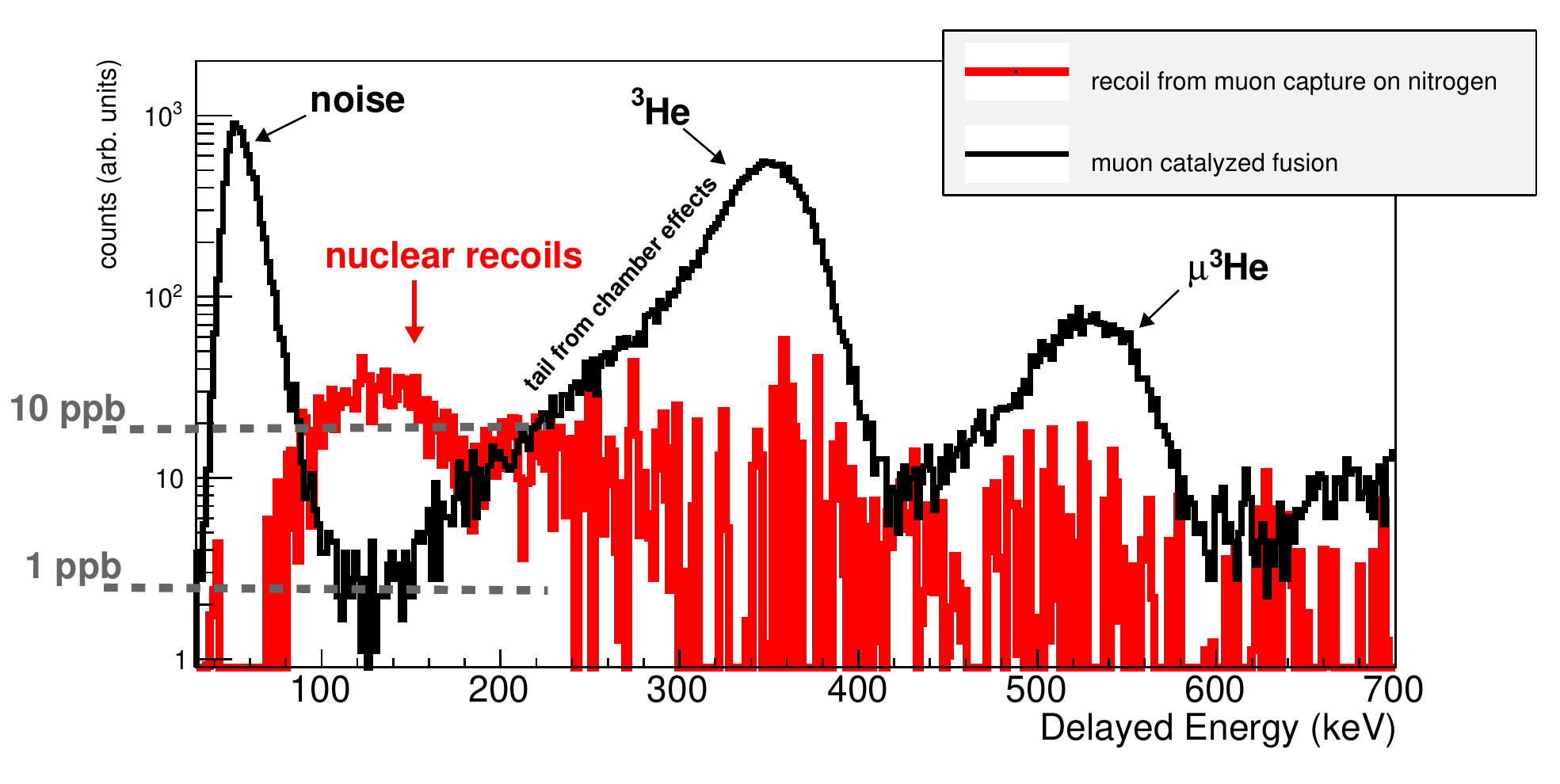}
\caption{Energy spectrum for delayed pulses following muon stop, collected 
with the new cryogenic preamplifiers in deuterium with N$_2$ content 
of $\sim$20~ppb, as confirmed by gas chromatography measurements.
The black spectrum includes only events that have an observed Michel electron,
which emphasizes the muon-catalyzed fusion signals.  The red spectrum,
marked ``nuclear recoils,'' has this background subtracted in order to 
isolate the impurity capture signal.}
\label{fig:beamSpectrum}
\end{figure}

The primary motivation for the development project
was to enable the MuSun TPC to be used as an in-situ impurity monitoring tool 
to measure chemical impurities in the deuterium at the part-per-billion scale.
An energy spectrum of delayed pulses after the muon stop is shown in 
Figure~\ref{fig:beamSpectrum}.  The deuterium gas was contaminated with 
$\sim$20~parts per billion of nitrogen at the time this spectrum was collected. 
The dramatically improved resolution in this spectrum may 
be seen by comparison with Figure~\ref{fig:R43He}; the gas density is also 
different, so the features are not at precisely the same energies, due to 
recombination effects.  The RMS width of the $^3$He fusion peak 
is 17~keV(D$_2$).  Crucially, the signal from the nuclear recoils following 
muon capture is clearly visible at an energy of 125~keV, separated from the 
noise peak at lower energies and from the fusion products at higher energies.
This measurement indicates that the new preamplifiers 
allow sensitivity to impurities at the 1~ppb level, providing an 
order-of-magnitude improvement over the room temperature preamplifiers.

\newpage 
\vskip 0.2in
\noindent
{\bf Acknowledgments}
\vskip 0.1in
\noindent
We gratefully acknowledge the work of all of our MuSun collaborators, 
and we thank them for permitting us to use the MuSun data in this publication.
We thank PSI for beam time and important experimental resources.
This work was supported in part by the National Science Foundation (grant 
number PHY-1206039) and the U.S. Department of Energy (grant number 
DE-FG02-97ER41020).  We thank Uwe Greife of the Colorado School of Mines and 
Bernhard Lauss of PSI for support in testing early prototypes.

\bibliographystyle{JHEP}
\bibliography{PreampJINST}

\end{document}